\documentclass[a4paper,9pt, twocolumn]{article}

\usepackage{graphicx} 
\usepackage[a4paper,margin=1.31cm, marginparwidth=0cm]{geometry}
\usepackage{float}
\usepackage{array}
\usepackage{graphicx}
\usepackage{ragged2e}
\usepackage{amsmath}
\usepackage{amssymb}
\usepackage{cleveref}
\usepackage{xcolor}
\usepackage{abstract}
\usepackage[backend=biber, sorting=none]{biblatex}
\usepackage[switch]{lineno}

\title{Unveiling individual and collective temporal patterns in the tanker shipping network}
\author{Kevin Teo\textsuperscript{1}, Naomi Arnold\textsuperscript{1}, Andrew Hone\textsuperscript{2}, Michael Coulon\textsuperscript{3}, \\Martin Ireland\textsuperscript{3},  Mauricio Santillana\textsuperscript{4}, Istv\'an Z. Kiss\thanks{istvan.kiss@nulondon.ac.uk} \textsuperscript{,1,5}}

\date{
\textsuperscript{1} Network Science Institute, Northeastern University London, London, E1W 1LP, United Kingdom\\
\textsuperscript{2} School of Mathematics, Statistics \& Acturial Science, Univeristy of Kent, Canterbury CT2 7NF, United Kingdom \\
\textsuperscript{3} AlphaOcean.ai Ltd, Preston Park House, South Road, Brighton, East Sussex, United Kingdom, BN1 6SB \\
\textsuperscript{4} Machine Intelligence Group for the Betterment of Health and the Environment, Network Science Institute, Northeastern University, Boston, MA, USA \\
\textsuperscript{5} Department of Mathematics, Northeastern University, Boston, MA 02115, USA \\
\today}

\begin{document}


\twocolumn[
\maketitle 
\begin{onecolabstract}

The global shipping network, which moves over 80\% of the world’s goods, is not only a vital backbone of the global economy but also one of the most polluting industries. Studying how this network operates is crucial for improving its efficiency and sustainability. While the transport of solid goods like packaged products and raw materials has been extensively researched, far less is known about the competitive trade of crude oil and petroleum, despite these commodities accounting for nearly 30\% of the market. Using 4 years of high-resolution data on oil tanker movements, we employ sequential motif mining and dynamic mode decomposition to uncover global spatio-temporal patterns in the movement of individual ships. Across all ship classes, we demonstrate that maximizing the proportion of time ships spend carrying cargo --a metric of efficiency-- is achieved through strategic diversification of routes and the effective use of intra-regional ports for trips without cargo. Moreover, we uncover a globally stable travel structure in the fleet, with pronounced seasonal variations linked to annual and semi-annual regional climate patterns and economic cycles. Our findings highlight the importance of integrating high-resolution data with innovative analysis methods not only to improve our understanding of the underlying dynamics of shipping patterns, but to design and evaluate strategies aimed at reducing their environmental impact.
\end{onecolabstract}
]


\noindent The global maritime network is a key infrastructure of the global supply chain, handling over 80\% of all goods by volume, and notably responsible for over 60\% of crude oil and petroleum product trades~\cite{UNCTAD_Review_Maritime_Transport_2017, EIA_World_Oil_Transit_Chokepoints}. Yet, its vast scale of operations results in significant environmental impact, with a major carbon footprint and a range of environmental pollution and ecological damage~\cite{rahim2016regulating, walker2019environmental, shi2023perspectives}.

Despite this, there is a relative lack of studies of maritime shipping networks, even within the domain of transportation networks~\cite{ducruet2020geography}.
Moreover, studies on oil tankers are comparatively sparser than those on container liners (ships moving solid and/or packaged goods). This is partially due to the fact that oil tankers are typically less predictable than container liners~\cite{kaluza2010complex}: they tend to not have fixed routes, and as a result they can compete for cargo. Consequently, the precise journey of a ship is strongly influenced by supply and demand trends, the trading strategies of charterers, and changing geopolitical relations. 
Additionally, obtaining data has been a challenge, as the shipping industry has traditionally been slower in digitalization and adopting data analytics tools compared to other transportation sectors~\cite{fruth2017digitization,tijan2021digital}. However, this trend began to shift due to several pressures, including regulatory restrictions, environmental mandates, and the need to optimize costs. As a result, the quantity and quality of data has been constantly improving. 

Benefiting from this improved data collection trend, we recently acquired a dataset of highly detailed timestamped ship movements across a four-year period pertaining to dirty oil transport. As our dependence on the maritime industry grows, a deeper and more comprehensive understanding of the interplay between economic and environmental factors - and how they shape decisions and behaviors from the scale of individual ships to entire fleets - becomes essential. Leveraging this comprehensive dataset, our study develops a principled understanding rooted in rigorous data analysis techniques. We uncover key emergent trends and patterns in shipping activity, providing actionable insights to promote and support practices that generate positive economic and environmental impacts.

Traditionally, maritime shipping activity has been modeled as a network of ports connected by shipping routes. 
This network, and the activity on it, has been studied to understand the different roles that various ports play (centrality), the ease with which one can navigate this network (connectivity), and it's robustness to technical or political disturbances~\cite{hu2009empirical, kaluza2010complex, kanrak2019maritime, ducruet2012maritime, ducruet2012worldwide, ducruet2020geography, liu2022structures}. These studies were carried out from various perspectives, including economic interests~\cite{liu2022structures, asgari2013network} and ecological interests~\cite{seebens2016predicting, seebens2013risk} . 
In particular, it was established that goods can be transported between any two ports in the world through this network with a small number of transit stops on average. Furthermore, most of the shipping routes were shown to pass through only a few key ports, identified in the network as ‘hubs’ that attract a lot of activity or `bridges' that connect different clusters of ports together \cite{ducruet2012maritime, zhang2022structuralhole}.   

Although these prior works examined static (time-aggregated) network-centric properties, our analysis delves deeper by explicitly accounting for the temporal and sequential order in the data. This is conducted on two key physical scales:
(i) individual ship behaviors, and (ii) regional fleet-wide trends. 
We capitalize on the full availability of temporal (past and present) information tagged to individual ships to study sequential motifs - short repeating patterns - within ship journey sequences, which was not possible with anonymized aggregated fleet-level itineraries. Building on this, we investigate whether ships with higher efficiency, as indicated by their laden-ballast ratio (a measure of the relative time spent carrying cargo during voyages), exhibit distinct traversal patterns compared to less efficient vessels, linking operational success (and potential carbon footprint) to movement behaviors.

Next, we shift our attention to the macro-scale, extracting periodic temporal patterns in regional shipping activity. Unlike other commonly studied complex networks such as social networks or online networks, the global maritime network is comparatively stable.
Changes typically occur as shifts in the frequency of trips, rather than additions or removals of new ports and/or routes. Limited by the relatively slow growth of the global fleet, supply and demand pressures induces a constant ebb and flow of shipping vessels, where an influx of ships into a region must be compensated elsewhere by an outflow of ships. As such, one cannot treat the sub-components of the system in isolation. One might therefore expect some level of synchronicity (or asynchronicity) between different regions, with potential links to wider economical or natural phenomena, such as the market and climate. 

Prior research on detecting periodicity in networks utilized network similarity measures to quantify the statistical difference between snapshots of the network at two different points in time~\cite{clauset2012persistence, masuda2019detecting, sugishita2021recurrence, andres2024detecting}. Such methods inevitably reduce the network into a single measure, potentially losing information on many features of the network.
To overcome this limitation, we employ a method known as Dynamic Mode Decomposition to extract periodic trends from multiple simultaneous time series signals, enabling more advanced temporal network-based analysis, as it avoids the need to reduce the entire network to a single value. We uncover pronounced seasonal signals in the regional shipping activity and temporal correlations between different regions, linking them to supply-demand dynamics in the energy and transportation sectors, which in turn are driven by seasonality in climate patterns.

\section*{Results}
\subsection*{Data} 
We obtained proprietary data from our partner company AlphaOcean from 2016 January to 2020 March covering voyages taken by $3081$ medium and large crude-oil and petroleum tankers. This consists of $458$ Panamax, $1156$ Aframax, $619$ Suezmax, and $848$ Very Large Crude Carrier (VLCC), listed in order of their size in deadweight tonnage. 
The dataset consists of a list of laden legs (single trips where the ship is loaded with cargo), detailing the source/destination location and their respective departure/arrival times, as well as an identifier of the ship that completed the leg. This enables the reconstruction of the entire historical journey of an individual ship, providing a unique opportunity to analyze sequential correlations in its trajectories by exploring how prior locations influence future destinations. While the locations provided are accurate up to the port-level, our analysis condenses this to a \textit{regional} level, combining $1337$ ports to $26$ regions. 
This is consistent with industry practices of pricing freight rate indices on a region-to-region basis, and also reflects the fact that most regions have large numbers of ports across which a cargo's exact discharge location may not be provided to the ship operator until quite late in the voyage.  Furthermore, ships occasionally make multiple consecutive discharges or pick up multiple consecutive loads, usually within the same region; these are merged in our dataset so as to not overestimate the number of chartered voyages by ship owners (see supplementary information for extra discussion). Consequently, the sequential record of a ship's journey alternates consecutively between loading and discharging regions. Self-loops, where a ship makes a trip within the same region, are allowed, though their prevalence is reduced by the merging of consecutive loading/discharging mentioned above. 

\begin{figure*}[ht!]
\centering
\includegraphics[width=1\linewidth]{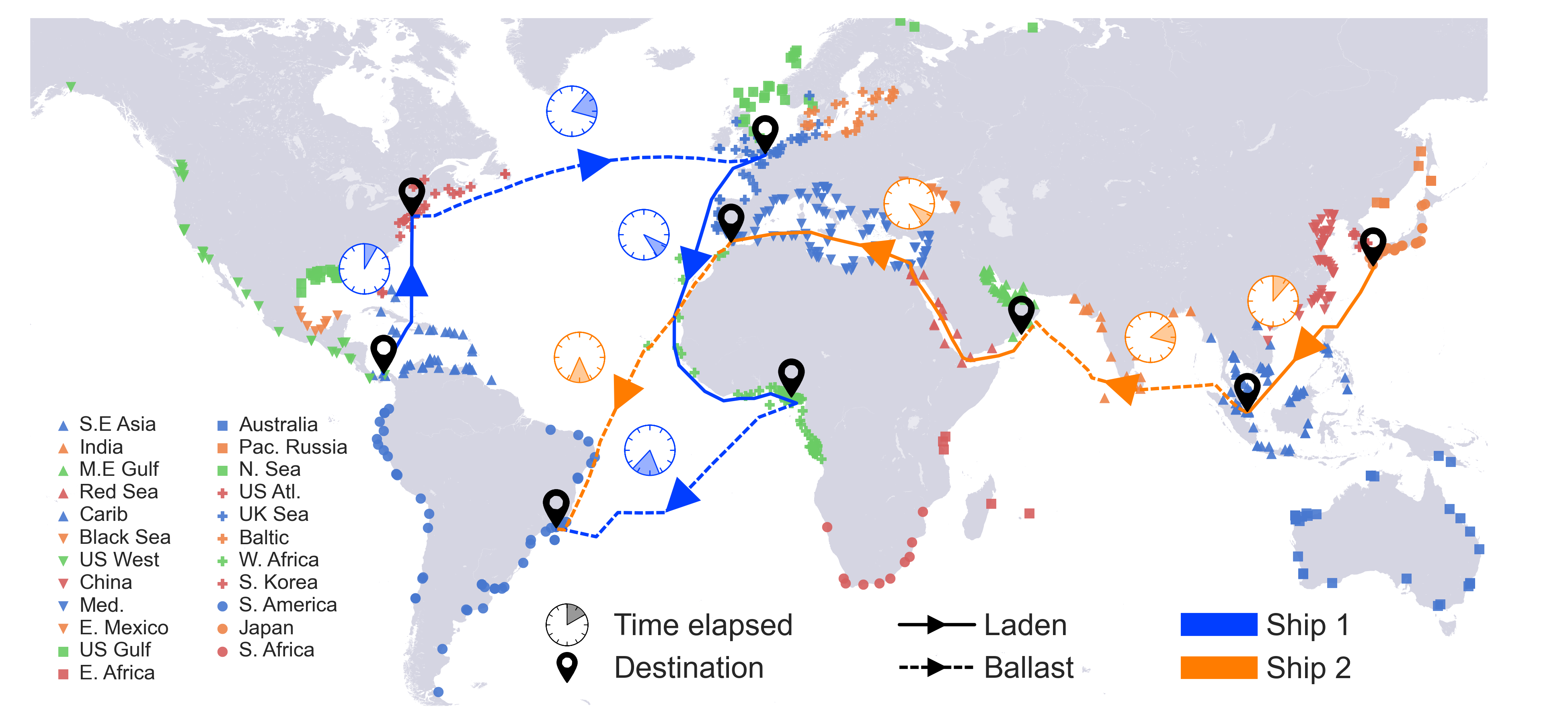}
\caption{\justifying 
    \textbf{Schematic representation of the journeys in time and space of two ships.} Ships travel along directed port-to-port edges with journey time indicated by clock markers. Individual ports in the dataset are also shown, with the regions they belong indicated by their marker shape and color.}
\label{fig:temporal-network}
\end{figure*}

Ballast legs (where a ship is not carrying cargo) are not explicitly recorded; instead, we infer their occurrence between two consecutive laden legs of the same ship: given that a particular ship performs two laden legs from regions $a\rightarrow b$ and then $c\rightarrow d$, this implies a ballast leg from $b\rightarrow c$. Note that, as the only timestamps recorded are for the departure and arrival time of each laden leg, the duration of ballast legs are consequently on average artificially longer than that of laden legs, as the additional amount of time spent on loading and discharging is not known. The data can be represented via an interval network, as shown in~\cref{fig:temporal-network} (detailed in Methods). In total, we recovered over $192,000$ shipping legs (both laden and ballast), split amongst $33,509$ Panamax, $94,840$ Aframax, $34,401$ Suezmax, and $29,418$ VLCC legs.

\subsection*{Ship performance indicators}
The ships we are concerned with typically do not travel on fixed routes, nor service a particular region. They are unconstrained and allowed to compete for cargo in different markets and regions. Their resulting movements cannot be modeled as random walks on top of a network of regions (or ports); instead, their movements are driven by financial and economic  considerations, such as travel times, bunkering and fuel costs, staffing, and potential earnings, to name a few. 
A further major factor influencing a ship's journey is the common industry practice of chartering single laden legs at a time, where the ballast leg required to reach the chartered pickup location is left up to the ship owner.  Since the voyage revenue for each ballast-laden pair is derived entirely from the laden leg, minimizing ballast legs is a key consideration for shipowners to ensure efficient ship management. 
The \textit{laden-ballast ratio}, which indicates the relative proportion of time spent laden during a ship's journey, arises as a natural measure for the efficiency of a ship. This measure reflects both the financial earnings associated with more time spent laden, and the cost savings associated with less time spent ballast, indicating an efficient usage of fuel, cargo space, and time. This captures a vital area of concern for the industry as key players shift towards more data-driven operations.

\begin{figure}[ht!]
    \centering
    \includegraphics[width=1\linewidth]
    {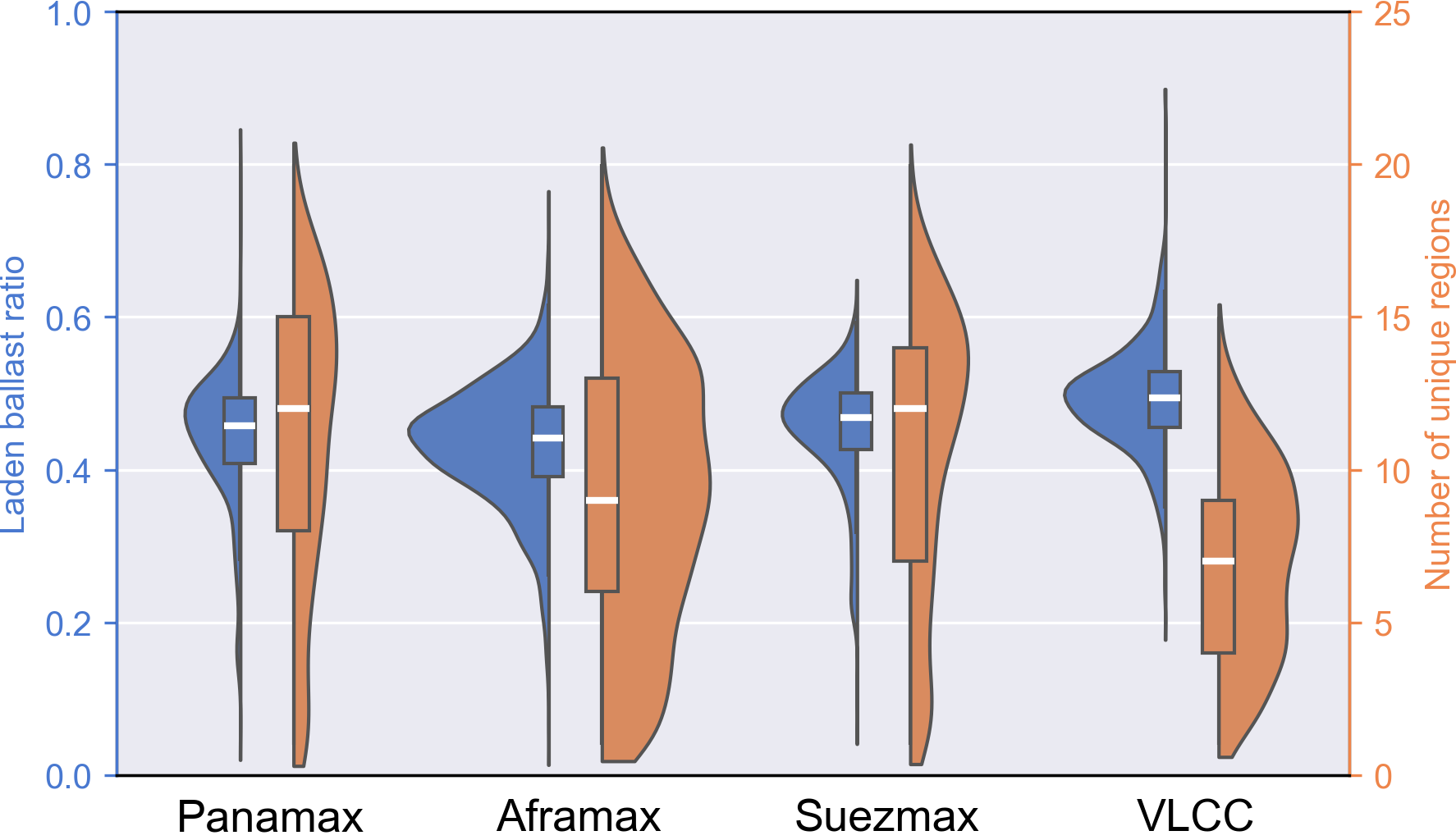}
    \caption{\textbf{Distributions of the laden-ballast ratio and the number of unique regions visited}. A split violin plot is shown for each ship class. Quartiles are shown by the corresponding box plots, with the median given in white. Width of each violin is proportional to the number of ships in that class, and the ship classes are ordered from left to right in increasing size (deadweight tonnage).
    }
    \label{fig:violin-plots}
\end{figure}

As shown in~\cref{fig:violin-plots}, all ship classes show similar median values in their laden-ballast ratio except for VLCCs, which show a marginally higher median value. 
The laden-ballast ratio distribution is consistently unimodal and symmetric, centered around a dominant mean value with longer tails at low values for all ship classes except VLCCs. 
It should be noted that there is no \textit{a priori} expectation as to what a typical average laden-ballast ratio should be.

While certain ports and regions dominate the maritime traffic flow, the ships themselves are not necessarily tied down to a particular region as a `home' or `base'. We investigate the freedom of ships to explore different markets and regions, and whether they may benefit from this.
To do so, we compare the distribution of the number of unique regions visited by a ship during its journey, shown in~\cref{fig:violin-plots}.
Here, we observe more variation, in contrast with the laden-ballast ratio. The classes Panamax and Suezmax show a similar average and distribution, with a median of 12 regions. Aframax, on the other hand, has a wider distribution of values with a median of 9 regions, likely due to their higher versatility and broader usage. VLCCs display a significantly different and narrower distribution to the rest, averaging at $7$ regions. This is likely due to VLCCs being the largest shipping class with a limited range of viable destinations, and therefore reserved for long-haul trips. 

At first glance, the variation in distribution shapes of the number of unique regions appears to be at odds with the laden-ballast ratio, where the distributions observed there are very similar. However, this variety suggests that there are indeed different underlying mechanisms behind the trajectory patterns of each ship type, while they are all attempting the same economic optimization task. 
Due to the nature and constraints of the system, this yields a similar laden-ballast ratio despite these differences in mechanistic behaviors. These minute differences nonetheless manifest themselves significantly, as these ships deal with enormous amounts of cargo and therefore considerable revenues with each voyage. Small percentage differences in laden-ballast ratio can translate to significant financial gains or losses, and may lead to different environmental impacts.

\subsection*{Extracting sequential motifs}
Despite the fact that the next region a ship will sail towards is influenced by many factors, studies have shown that its next move can be modeled statistically using higher-order Markov chains, where future locations are influenced not only by their current location, but also by their past \cite{chawla2016representing, teo2024performance}. Knowledge of the two most recent locations (including the current location) has been shown to significantly enhance predictions of the next location. Interestingly, this applies to all ship classes. This may potentially be explained by the different types of trip taken in relation to the functions of different ports in regions. For example, a ballast journey into a net-exporting region suggests potential future destinations --perhaps to a net-importing or politically aligned region-- that may differ from a laden journey into the same region. We survey the journeys of ships through this lens of higher-order sequential correlations. In this case, differences between ship classes are minimal, and thus we aggregate across all ship classes. 

\begin{figure*}[htbp]
    \centering
    \includegraphics[width=1\linewidth]{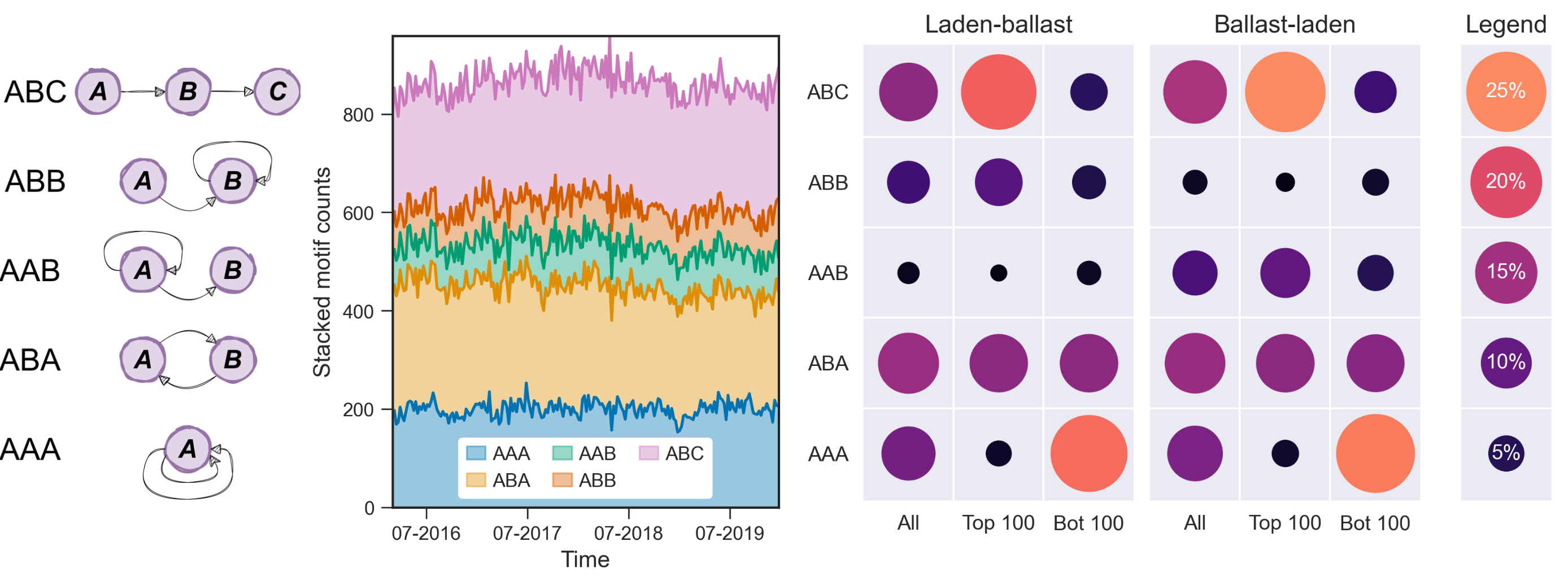}
    \caption{\justifying \textbf{Observed motif distributions of ship movements.} (Left) Diagrams of $2$-hop motifs. (Center) Stacked plot of motif counts over time in $1$ week windows. Counts indicate the number of motifs that \textit{started} within each $1$ week window, as these $2$-hop trips are not instantaneous and vary in their duration. (Right) Motifs of the top 100 and bottom 100 ships compared to the whole fleet.}
    \label{fig:motifs}
\end{figure*}

We first consider the prevalence of small \textit{sequential motifs} --simply referred to as \textit{motifs} from here on-- which represent recurring traversal patterns~\cite{milo2002network,kovanen2011temporal,paranjape2017motifs,larock2022sequential}. In graph theory, a sequential motif of length $l$ is a small directed subgraph which can be traced out end-to-end by a single walk of length $l$ (i.e. a walk of $l$ edges) through a set of nodes. Motifs are described in terms of the walk traced out by them, e.g. ``ABAB'' is a motif of length $3$ that describes a walk which starts from $A$, then hops to $B$, then back to $A$, and then back to $B$. Crucially, the specific labels of the nodes are not important; instead, the topological relations between nodes are what matter in identifying motifs. For example, a journey sequence \textit{(China, Japan, China)} and \textit{(Japan, China, Japan)} both map onto the same motif $ABA$. Furthermore, journeys on a different set of nodes, such as \textit{(UK, Mediterranean, UK)}, also map onto the motif $ABA$. Up to a relabeling of nodes, there are five possible motifs of length $2$, which we call $AAA$, $ABA$, $AAB$, $ABB$, and $ABC$, shown in~\cref{fig:motifs}. Recall that in this context, a self-loop refers to a leg with a source and destination in the same region. Using~\cref{fig:temporal-network} as an example, the journey of ship $1$ can be recorded in regional terms as \textit{(Caribbean, US Atlantic Coast, UK Continent, West Africa, South America)}. Breaking this down into $2$-hop motifs, this amounts to $3$ $ABC$ motifs.

We begin by considering whether the frequency of these $2$-hop motifs changes significantly over time. This may indicate a change in the behavior of ships in the network. We look at the absolute counts of different motifs over time, shown in~\cref{fig:motifs}.
The general trend shows that despite the random fluctuations in counts, the relative prevalence of each motif does not vary significantly over time, 
with $ABC$ and $ABA$ equally dominating, while $AAB$ and $ABB$ are equally weak, suggesting that ships do not vary their behavior or strategy substantially. This allows us to aggregate across the whole duration of the data without losing critical information. Moreover, this trend holds when looking across different ship types as well. 

We turn to look specifically at the composition of motifs associated with highly efficient ships. To investigate this, we count the motifs of the top 100 and bottom 100 performing ships as indicated by their laden-ballast ratio, and compare them to each other and to the population average, as shown in ~\cref{fig:motifs}. Firstly, there is a remarkable difference in relative counts of $AAA$ and $ABC$ motifs, with each being overwhelmingly preferred by the bottom and top performing ships, respectively. This indicates that the top-performing ships prefer to explore different regions and avoid staying within the same region. Moreover, as seen by the self-loops in $AAB$ and $ABB$ motifs, top-performing ships use intra-region trips as ballast legs, which suggests that the ideal reason for intra-region travel is to pick up nearby cargo. 

Taking a closer look at the results for VLCC, there is a seeming contradiction where VLCCs both exhibit a higher laden-ballast ratio, and have a high propensity for the $ABC$ motif, while having a lower average number of unique regions visited. This can be reconciled by noting that, while VLCCs do indeed operate only within a smaller total number of regions, they move unpredictably within these few regions. 
Furthermore, by focusing on longer voyages, they benefit from a relatively smaller proportion of days in port.

The key insight from our motif results can be highlighted by further investigating trips involving the Middle East (see supplementary information).
The most dominant patterns across all ships involve round trips going towards or from the Middle East, where ships typically load, then unload their cargo elsewhere before returning to the Middle East.
However, this pattern does not hold for the top 100 ships. In other words, the top performing ships favor more exploration over a steady business with a major exporter, indicating the potential value of chasing the market.
From this we conclude that the diversification of targets is a key component in a successful shipping strategy, as well as capitalizing upon intra-region ports for shorter ballast legs. 
Moreover, we estimate that up to 270 tons of fuel per month could be redirected from ballast (empty) legs to laden (fully loaded) legs in a medium-sized (and continuously operated) vessel following better routing strategies \cite{ADLAND2020optimalship}. 
This represents approximately 25\% of monthly fuel costs, which could not only lead to increased revenue from completing more (or longer) laden legs within the same time frame, but also mitigate excess emissions from sailing empty. If all of the bottom performing ships improved their routing strategies, these multi-vessel efforts could lead to sizable improvements in ship utilization rates, potentially lowering the overall carbon footprint by reducing excess ballast trips.

\subsection*{Identifying periodicity in shipping activity}
We now shift our focus to macro-scale features of the network, looking for emergent phenomena arising out of coordination and competition from these ships. Variation in the market creates pressure differentials, and thereby  opens up new opportunities for ships to benefit from. These variations may be unpredictable, potentially due to political factors or plain accidents. However, they may also be predictable, due to cyclical physical or social phenomena. These range from climate patterns~\cite{stopa2014periodicity} to socioeconomic patterns in production, operation, or consumption. In order to capture these dynamics, we utilize an interval graph, detailed in Materials and Methods, where regions are represented as nodes. We analyze the evolutionary dynamics of node-level properties, specifically the total number of incoming ships (both laden and ballast), referred to as the node \textit{in-degree}. While the regional shipping network structure remains relatively stable for a complex system, there are variations in the inter-region dynamics, often in response to external forces and influences. 

A crucial step in detecting periodicity in signals is to first identify and remove non-periodic trends, such as linear growth and/or decay, as they can overwhelm the periodic signals of interest. In the context of the shipping network for example,
China is a growing region, with an increasing year-on-year demand for crude oil and petroleum products  over this time period \cite{EIAchina, sinton2000goes} (see supplementary information for details). However, such changes are constrained by the logistics of the system; the production and transport of crude oil and petroleum products are limited by the number of oil producers, and the number of ships in the system. We can therefore infer that growth or decay in this context are approximately linear, not exponential. The general step then is to assume, for each region, the true signal $f(t)$ takes on the functional form 
\begin{equation}
    f(t) = mt + c + y(t), \qquad y(t) = s(t) + \varepsilon(t)
\end{equation}
where $mt + c$ is a straight-line fit, and $y(t)$ is a periodic function containing a purely periodic term $s(t)$ and some random noise $\varepsilon(t)$ \cite{hylleberg2014seasonality}. The sum of terms $mt + c$ acts as a linear background growth or decay, which can be determined through standard linear regression. Therefore, as part of the pre-processing of the data, we remove this linear term to keep only $y(t) = s(t) + \varepsilon(t)$.

We employ the Dynamic Mode Decomposition (DMD) method \cite{schmid2010dmd}, specifically bagging-optimized DMD \cite{Askham2018.bib, sashidhar2022baggingdmd}, implemented in \verb|Python| with the \verb|PyDMD| package \cite{demo2018pydmd, ichinaga2024pydmd}. Given a vectorized time series $\mathbf{y}(t) = (y_1 (t), y_2(t), \ldots, y_N(t))$ over $N$ regions, the $R$ rank DMD method finds a complex-valued reconstruction 
\begin{equation}
\begin{aligned}
    \mathbf{\hat y}(t) = \sum_{j=1}^R \mathbf{\Phi}_j \exp(i\omega_j t) 
\end{aligned}
\label{eq:dmd-exp}
\end{equation}
where $\mathbf{\Phi}_j = (\phi_{1,j}, \phi_{2,j},\ldots, \phi_{N,j}) \in \mathbb{C}^N $ is a vector of  complex magnitudes and $\omega_j \in \mathbb{R}$ is the frequency of the $j^{th}$ mode. The $\mathbf{\Phi}_j$ and $\omega_j$ terms are parameters to be fitted. In this case, the real part  of $\mathbf{\hat y}$ is the best-fitted reconstruction of the data. 

We can decompose the terms inside the right-hand side of~\cref{eq:dmd-exp} into more intuitive forms. In particular, the real-component of the $n^{th}$ region of the $j^{th}$ mode can be written as  
\begin{equation}
    \Re[\phi_{n,j} \exp(i\omega_j t)] = A_{n,j}\cos{(2\pi t/\tau_j + \theta_{n,j})},
    \label{eq:dmd-real}
\end{equation}
where $A_{n,j}$ and $\theta_{n,j}$ are obtained from the polar form of the complex number $\phi_{n,j} = A_{n,j} \exp(i\theta_{n,j})$, and $\tau_j = 2\pi/\omega_j$ is the period. In our analysis, we selected $6$ distinct modes for the reconstruction. The reconstruction results for the top $7$ regions from the DMD method are shown in~\cref{fig:dmd-recon}, and the results for $3$ modes are shown in~\cref{fig:dmd-map} (see supplementary information for reconstruction of all regions and all modes).

\begin{figure}[t!]
    \centering
        \includegraphics[width=1\linewidth]{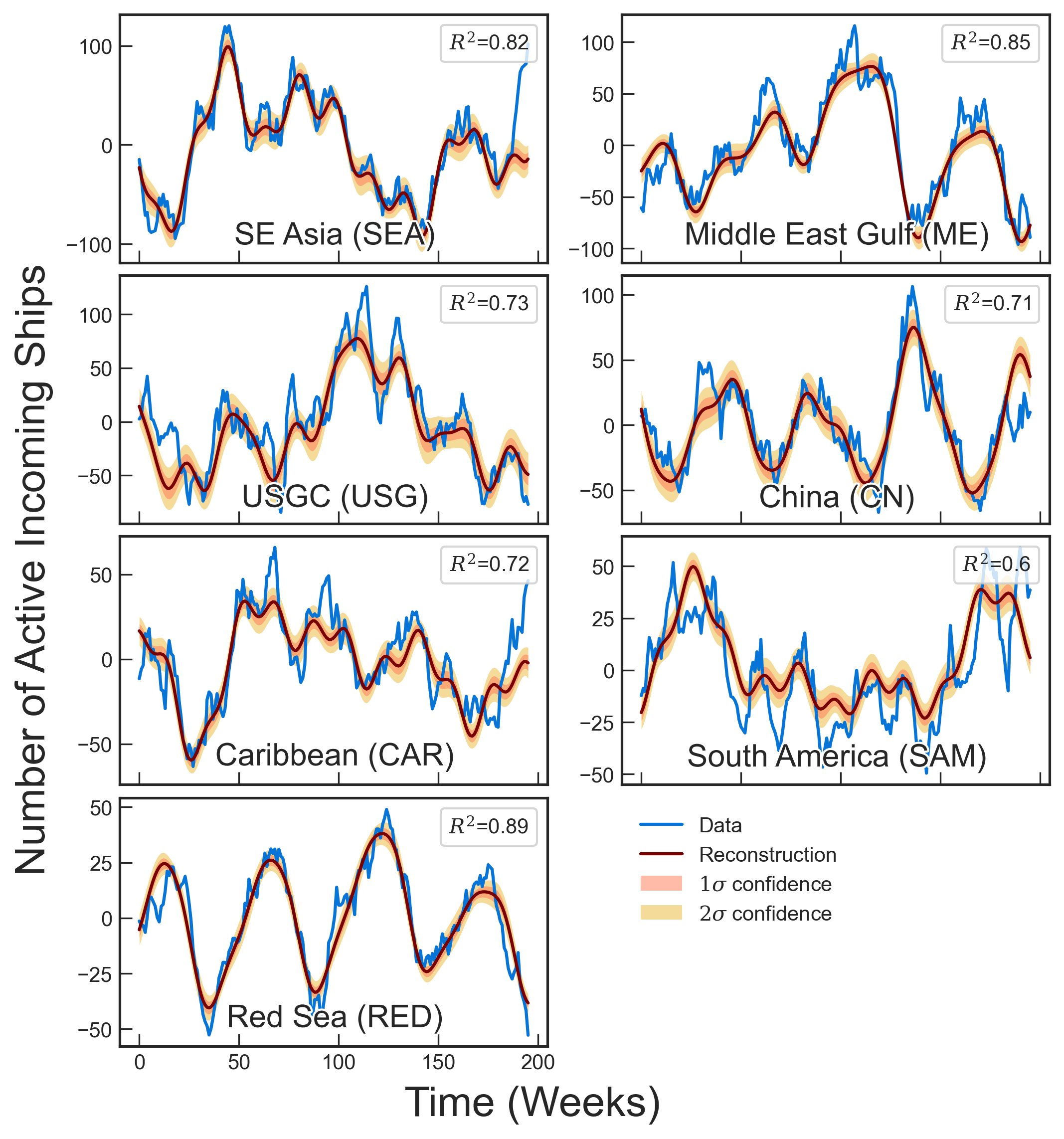}
        \caption{
        \justifying \textbf{Top 7 major regions' in-degree time series.} The preprocessed data (blue) and the DMD reconstructed signal (brown) is shown alongside a 1 and 2 $\sigma$ confidence interval.}
        \label{fig:dmd-recon}
    %
\end{figure}

\begin{figure*}[h!t]
    \centering
    \includegraphics[width=1\linewidth]{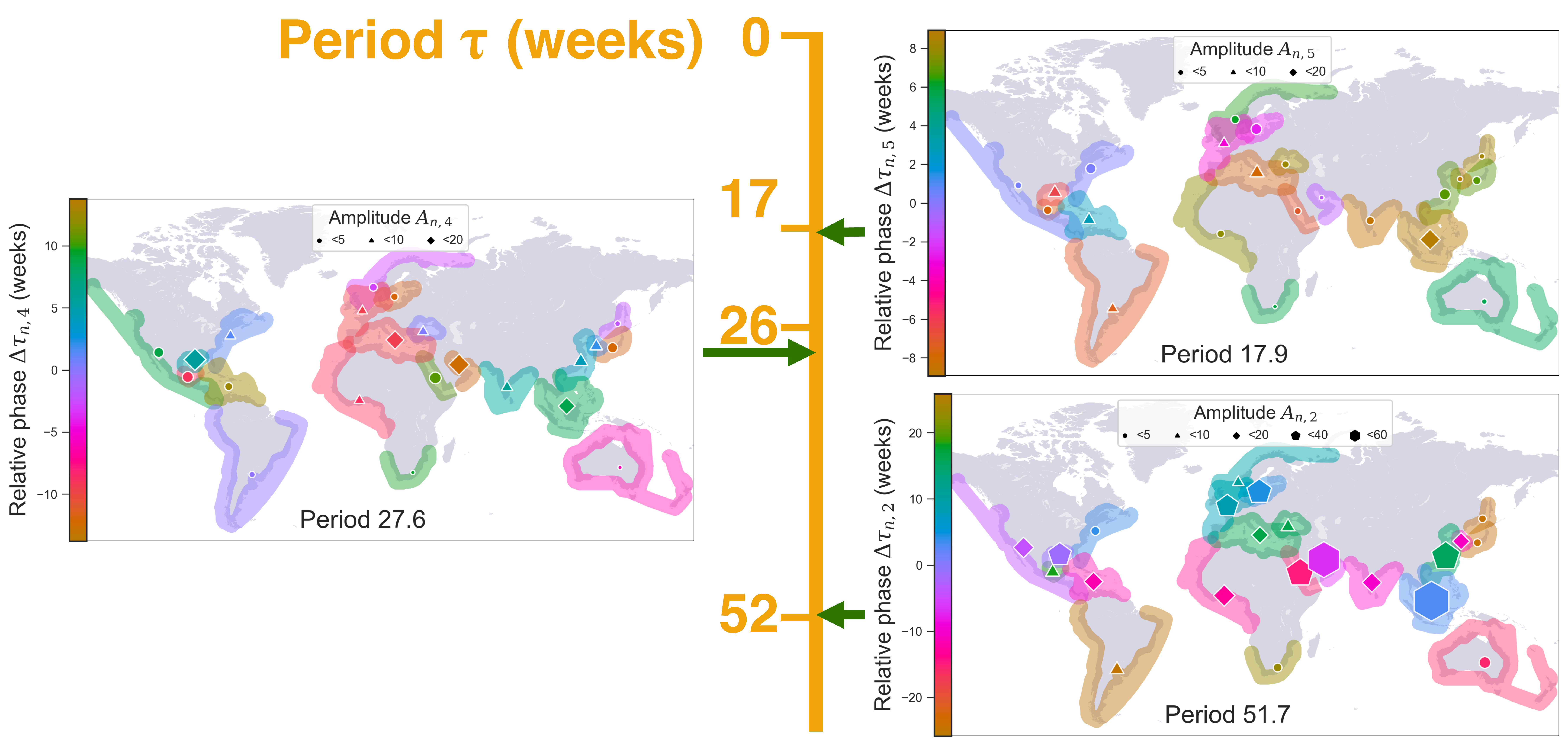}
    \caption{\justifying \textbf{DMD mode amplitudes and phases.} Visualization of modes $A_{n,j}\cos(2\pi t/\tau_j +\Delta\tau_{n,j} \times \pi/\tau_{n,j})$ with periods $\tau_j$ for $j=2, 4, 5$ with each region indexed by $n$. Each mode is shown by a world map in which each region is represented by a single marker, with a shaded area that  covers the locations of its ports. The marker size and shape shows the amplitude $A_{n, j}$, while the hue of the marker and shaded area shows the relative phase $\Delta \tau_{n,j} = \theta_{n,j}\times \tau_{j}/\pi$. Only $3$ out of the total $6$ modes are shown here (see supplementary information for all modes).} 
    \label{fig:dmd-map}
\end{figure*}

First, we evaluate the DMD outcome by looking at the reconstruction of the time series data on a regional level, measured on a weekly interval with a $6$-week rolling window. In~\cref{fig:dmd-recon} we show the reconstruction of the top $7$ regions, in terms of their variation amplitudes. The goodness-of-fit was measured using the $R^2$ coefficient of determination statistic. The $R^2$ statistic lies in the range $[0,1]$, with $0$ being no better than a flat straight-line guess and $1$ being a perfect match with the data. We can see that among the top regions, most have an $R^2$ value $>0.7$. Of the remaining regions, most of them have an $R^2$ value of $>0.5$, which shows that the fitted periodic trend accounts for more than $50\%$ of the observed variance, with only 4 regions clearly underperforming with the lowest $R^2=0.43$ (see supplementary information for the reconstruction of all regions). We can clearly see that the DMD method is recovering significant patterns in the data. In order to understand what these patterns are, we need to examine the parameters: the eigenmodes and eigenvalues.

The interpretation of the DMD outcome can be split into two main parts. Recall from~\cref{eq:dmd-real} that there are three key groups of parameters: the periods $\tau_{j}$, amplitudes $A_{n,j}$, and phases $\theta_{n,j}$. Each mode corresponds uniquely to a period $\tau_{j}$, which indicates the duration of a single cycle. Therefore, it is often convenient to refer to each mode by its period instead, e.g. mode $j=2$ is the mode with a period of $51.7$ weeks.  Out of the $6$ modes, we highlight $3$ specific modes and discuss their trends. These are the modes of $\tau_5 = 17.9$ weeks, $\tau_4 = 27.6$ weeks, and $\tau_2 = 51.7$ weeks.

For each mode, we take into account the amplitudes $A_{n,j}$, which indicate which regions (and modes) are associated with high variability/seasonality. A larger amplitude signifies a more pronounced seasonal cycle. 
The mode of $\tau_5=17.9$ weeks, or $\sim4$ months, shows heightened variability in the regions of South-East Asia, Mediterranean, US gulf coast, EC Mexico and UK Continent. Surprisingly, we observe little variability in the Red Sea, Middle East Gulf and China. 
Meanwhile, the mode of $\tau_4=27.6$ weeks, or $\sim6$ months, reveals increased variability in many regions including the US, UK Continent, Mediterranean, West Africa, Middle East Gulf, India, South-East Asia, and China. This semi-annual cycle is potentially linked with various factors, such as the consumption patterns of the industrial, commercial, and residential sectors, which are secondary drivers of demand. Additionally, while petroleum is not the main source of energy generation, it is still used as a supplementary fuel and usually experiences increased consumption in summer \textit{and} winter as extreme temperatures in either season drives up the usage of electrical heating and cooling. 

The mode of $\tau_2 = 51.7$ weeks, or $\sim 1$ year, is the largest amplitude mode, and therefore is also the mode that is most important to the overall dynamics.
Furthermore, this mode is the one that is most spread out geographically, with many regions around the world participating significantly in this cycle. We also observe well-known key regional players in the industry with South East Asia, Middle East Gulf, China, and Red Sea leading this pattern. This is the most expected periodic pattern, as the natural seasons shape physical and socio-economic patterns. There are several plausible direct factors that affect the movement of ships, in particular the economic supply and demand of goods, driven primarily by the transportation sector, with secondary influences from the energy sector as well. These sectors are in turn linked to seasonality in the weather, where, for example, many Northern regions' transportation fuel consumption increases in the winter (with the notable exception of the US, where it increases in the summer). 

However, the amplitude information alone is not enough to fully understand the relationships between different regions in a given mode, specifically around whether they are synchronized or not. This information is captured in the phase $\theta_{n,j}$, which indicates the phase shift in the $j^{th}$ mode of the $n^{th}$ region at time $t=0$. In other words, it shows where in a sinusoidal cycle each region is compared to each other at a given time. 

Looking at the modes of $17.9$ and $27.6$ weeks, we observe a time-lag of $6$ and $8$ weeks respectively between the Middle East and South-East Asia. 
Many of the most common repeated routes of ships occur between these two regions, and the resulting time-delay could be attributed to travel times between the two.  

For the mode of $51.7$ weeks, the regions around Europe and East Asia are approximately within a quarter phase of each other, with the largest outlier within this group being South-East Asia. This suggests that they are approximately in sync with each other, with their cycles matching in the range of $\pm 10$ weeks. This synchronicity cannot be explained by a time-delay induced by the travel times of ships between regions. Instead, these cycles are aligned due to changes in demand as the climate varies due to the natural seasons, where the influence is global. This group of regions consists of high-import regions in the Northern hemisphere, with low production capacities compared to their demand. 
In contrast, approximately anti-phased with these regions are the Caribbean, West Africa, Red Sea, India and Australia, running the band of tropical regions (as well as Australia). These regions do not experience the same kind of weather patterns as the Northern regions; and as such, the surging demand in the North is compensated by ships moving away from the tropics and Australia. 

\section*{Discussion}
The global tanker shipping network, despite its enormous economic and ecological impact, remains poorly studied, even within the broader field of maritime networks. Unlike the more commonly studied container liners, which operate on long-term schedules along fixed routes, tankers typically operate based on short-term chartered contracts, introducing additional complexity.
This study showcases that utilizing high resolution data we can distill important trends of the tanker network by explicitly treating the sequential and temporal elements of the system both at  the level of individual ships and at the regional level. 
 
Our ship-level analysis, by inspecting the variability in traversal patterns of individual ships and comparing the prevalence of motifs, revealed significant differences between more and less efficient ships. 
The most efficient ships are more varied in their travels and strongly abstain from staying within the same region, indicated by many $ABC$ and few $AAA$ motifs, while the inverse is true for the least efficient ships. Furthermore, the most efficient ships utilize intra-region trips almost exclusively for ballast journeys, minimizing their ballast time. 
This points to higher efficiency ships being more opportunistic, taking advantage of a strategy of local optimization instead of long-term stability, by targeting fixtures in regions with attractive freight rates at that time.

Shifting our attention to the regional scale, our application of the DMD method to the time series signals of regional shipping activities (node degrees in the network) detected significant periodicity at annual and semi-annual timescales. The most dominant mode corresponds to a cycle of $1$ year, likely reflecting the major seasonal phenomena driven by the social and economic supply and demand responses to seasonal climate patterns~\cite{hunt2003seasonalityUKJP,hao2020seasonalvehicle,rosado2021seasonalityportugal}. Similar cycles are observed in the transportation sector, the primary consumer of crude oil and petroleum products, potentially explained by the preferences for driving during colder months of the year. A smaller but still significant consumer is the energy sector, which experiences increased demand in energy production during both hotter and colder months in certain regions, leading to a $6$-month cycle, providing a link to the weaker $27.6$ week cycle observed in the fleet shipping activity. Other modes observed may be simply fitting corrections to random fluctuations, or higher-order correction terms for non-sinusoidal signals, such as saw-tooth signals.

While we focus on region-level (node) statistics in this study, as they provide a sufficiently high density of data over time, our methods are not constrained to node degrees. With sufficient data, it would be worthwhile to consider the periodicity of other network objects or properties, such as the dynamical evolution of edge weights, or on a more fine-grained network, such as the network of ports themselves. Furthermore, access to more specific measures such as revenue generated or fuel spent on voyages would be a substantial improvement over the laden-ballast ratio in the evaluation of these trends. 

Unpacking these ship- and region-centric trends may allow individual ship owners, port authorities, and regional authorities to better understand the expected dynamics of the system, as well as providing tools to aid in analysis and forecasting. Promoting increasing diversification for individual ships can help to reduce the vulnerability of the network and its dependence on a few major choke points, providing redundancy and resilience to sudden shocks. Furthermore, recognizing periods of low activity may offer an opportunity for implementing technological and infrastructural changes that support the industry's transition towards greater sustainability with minimal impact on the overall connectivity and robustness of the shipping network~\cite{UNCTAD_Review_Maritime_Transport_2017}. Finally, although our focus here has been on the tanker shipping network, our methodology can be adapted to suit a range of temporal systems with walk-embedded processes, in particular transportation networks.

\section*{Methods} \label{sec:methods}
\subsection*{Temporal Network}

To account for the dynamical evolution of the system, we use a temporal network representation \cite{holme2012temporal, masuda2016guide}, known as an interval graph or a stream graph. A schematic of this is shown in~\cref{fig:temporal-network}. Regions are represented as nodes, and every leg in the dataset corresponds to a directed temporal edge from the source node to the target node, and is considered active between the departure time and arrival time. Additionally, we also assign a `status' attribute to indicate whether the edge was laden or ballast. Finally, every edge is assigned to a layer associated with an individual ship. This allows us to query the network, between some time interval $[t_1, t_2)$ as to how many edges belonging to node $v$ are active \textit{across all layers}. This equates to measuring how many ships were entering or leaving region $v$ between times $t_1$ and $t_2$. This interval graph is implemented in \verb|Python| using the \verb|Raphtory| package \cite{Steer2024}.


\section*{Data availability}
Due to legal restrictions, the data used to generate the main results in this article will only be available by contacting the author Kevin Teo.

\printbibliography

\section*{Acknowledgments}
Kevin Teo acknowledges the PhD studentship support from Northeastern University.

\section*{Author contributions}
K.T, N.A, M.C, M.I, I.Z.K designed research; K.T, N.A performed research; M.S. contributed analytic tools; K.T, N.A, A.H, M.C, M.S, I.Z.K analyzed data; K.T, N.A, A.H, M.C, M.S, I.Z.K wrote the paper.


\section*{Additional information}

\subsection*{Correspondence}
Correspondence should be addressed to Kevin Teo or Istv\'an Kiss. Materials request should be addressed to Kevin Teo.

\end{document}